# A Study of the Circular Pursuit Dynamics using Bifurcation Theoretic Computational Approach

Kavita Shekhawat[1] and Nandan K Sinha[2]

Department of Aerospace Engineering, IIT Madras, Chennai 600036, India.

**Abstract**: A circular pursuit guidance problem involving pursuer-target engagement is studied in this paper using a bifurcation theory based numerical approach. While target is modelled as a point mass moving around in a circle with certain velocity, pursuer's dynamics is driven by the relative position and orientation with respect to the target. A planar case is currently considered. A mathematical model representing the engagement scenario is derived and two cases are presented, one without and the other with a basic model for pursuer speed dynamics accounting for limitations imposed by available force. Analytical and simulation results are presented to elucidate the novel approach. Advantages of using this approach for arriving at laws for pursuer-target engagement are highlighted.

## 1. Introduction

Pursuit problems involving leader-follower engagement in various scenarios are of immense interests in many fields of science and engineering. In aerospace engineering, scenarios involving an aircraft (pursuer) following either a moving target (another aerial vehicle) or a stationary point, are very common. There is huge literature on development of guidance laws around this problem, most modelling it as two body kinematic problem [1-4]. Due to complexity of the dynamic models, often nonlinear, theoretical development of guidance laws including dynamics of pursuer or target becomes infeasible. In very limited cases, for example in Ref. [5], some model for the pursuer dynamics is used. Guidance laws so developed usually need to be verified via software-in-the-loop (SIL) simulations using six-degree-of-freedom vehicle model. In a first, circular pursuit problem is attempted in this work from dynamical system perspective, using a numerical approach based on bifurcation analysis and continuation methods. The approach is purely computational and allows inclusion of nonlinear system models.

Bifurcation analysis and continuation theory methodology (BACTM) was primarily introduced as a tool for analysis and prediction of nonlinear behaviour of highly complex aircraft models. The approach involved computing steady states of aircraft model as function of a control parameter of aircraft using a numerical continuation algorithm, that also computed stability and multiple branches of steady states beyond loss of stability at critical value of the varying control parameter. Thus, gathering useful information, such as, existence of high angle-of-attack dynamics and how an aircraft may get into one, and reproduction of known phenomena like wing rock, spin, spiral departure, roll-coupled motions, via analysis of aircraft mathematical models became a standard practice in aircraft industry [6]. A major development in this field was development of an extended bifurcation analysis technique for analysis of aircraft flight dynamics in constrained conditions which required variation of multiple controls at the same time [7]. Many applications of bifurcation analysis methods in aircraft flight dynamics and control are excellently summarized in some reviews in Refs. [6,8,9]. Further, a recent noteworthy development utilizing numerical continuation technique has been in computing nonlinear frequency response for aircraft model at an operating condition of interest [10].

Some works using bifurcation as an aid in guidance law design are presented in [3-5], however, they are limited in the sense of restricting the theoretical development of guidance laws to simpler models and approximations based on assumptions. In his work [5], Paupolias studied loss of stability resulting in bifurcation thus affecting guidance law for autopiloting of marine vehicles. In recent works, bifurcations of look-angle [3] and line-of-sight [4] have been assumed apriori in simplistic forms forming the basis for theoretical development of guidance laws. In this work, a purely computational approach is adopted to compute guidance law in a circular pursuit making use of numerical continuation algorithm and integrated models of pursuer and target.

Paper is organized as follows: Section 2 presents a description of the problem preceded by kinematic model for an engagement scenario. In Section 3, a dynamical system model for the circular pursuit problem is derived. Further, a coupled model including pursuer dynamics is proposed in Section 4. Results are presented in Section 5 with discussions. Section 6 concludes the paper highlighting importance of this approach and future scope of work.

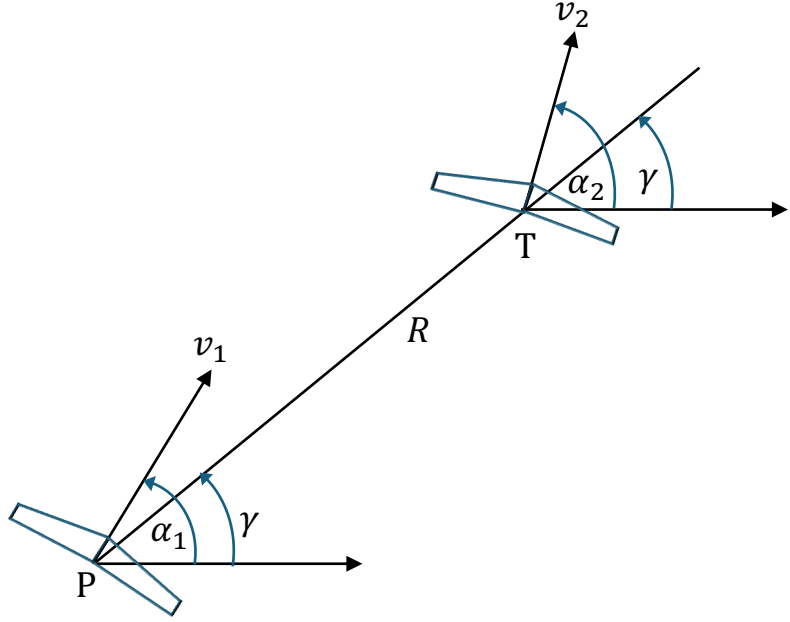

**Figure 1**: Target and follower engagement scenario.

## 2. Derivation of the mathematical model

A generic co-planar pursuit scenario is presented in the Fig. 1 of that of pursuer (P) and a target (T). Pursuer is moving at velocity $v_1$ at an angle $\alpha_1$ from local horizon chasing a target moving velocity $v_2$ at angle $\alpha_2$ from the horizontal axis. $R$ represents the distance between the pursuer and the target at any time with the straight line joining the two, LOS (Line-of-Sight), making an angle $\gamma$ with the horizontal axis. Kinematic equations representing above scenario (in the direction of increasing $R$ and perpendicular to LOS) are given by

$$\begin{aligned}\dot{R} &= v_2 \cos(\alpha_2 - \gamma) - v_1 \cos(\alpha_1 - \gamma) \\ R\dot{\gamma} &= v_2 \sin(\alpha_2 - \gamma) - v_1 \sin(\alpha_1 - \gamma)\end{aligned} \quad (1)$$

In Eq. (1), dot over variables represents time derivative. A special case of this engagement scenario is a 'circular pursuit' shown in Fig. 2. The 'Circular pursuit' described as a bifurcation problem taken from Ref. [11] is explained below.

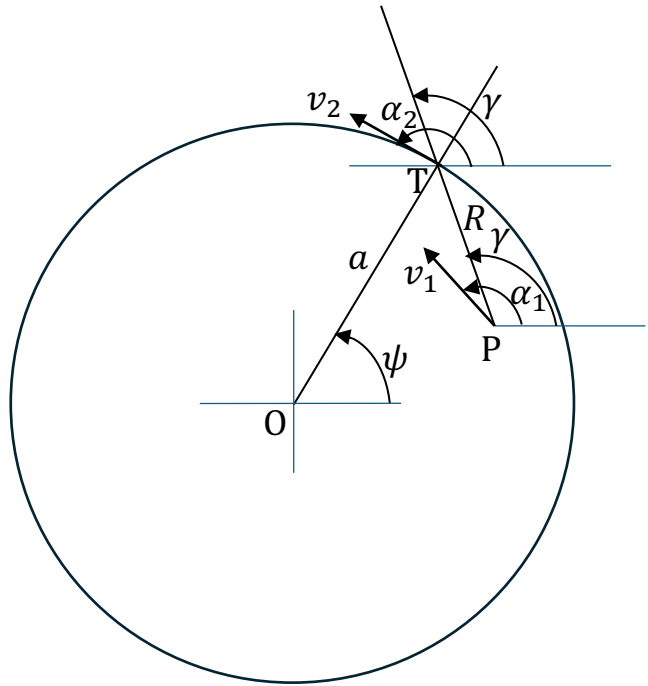

**Figure 2**: Schematic of a circular pursuit.

**Problem description (verbatim from Ref. [11], Exercise 7.1.9)**: A dog at the center of a circular pond sees a duck swimming along the edge. The dog chases the duck by always swimming straight toward it. In other words, the dog's velocity vector always lies along the line connecting it to the duck. Meanwhile, the duck takes evasive action by swimming around the circumference as fast as it can, always moving counterclockwise. (a) Assuming the pond has unit radius and both animals swim at the same constant speed, derive a pair of differential equations for the path of the dog. Analyze the system dynamics. Does the dog ever catch the duck? Can this problem be solved explicitly? (b) Parameterize the equation with a factor relating ratio of dog's speed to the duck's speed and study its parameterized dynamics.

In order to address the problem above, as a special case of the engagement scenario in Fig. 1, target is considered to be moving in a circle of radius $a$ with angular speed $\omega$ in counterclockwise direction as shown in the schematic in Fig. 2.

From Fig. 2, the following can be inferred,

$\dot{\psi} = \omega$; $\psi = \gamma - \phi$; $\alpha_2 - \gamma + \phi = \frac{\pi}{2}$; and $\alpha_1 - \gamma = 0$ for pursuer always pointing towards the target. $\phi = \angle OTP$. Substituting the above equality relations in Eq. (1) one can arrive at,

$$\dot{R} = v_2 \sin\phi - v_1 \tag{2}$$

Tangential speed of the target, $v_2 = a\omega$, which is fixed for given fixed values of $a$ and $\omega$. For ease of analysis, the above equation is non-dimensionalized with respect to angular distance covered by the target, $\psi = \omega t$, and a non-dimensionalized distance, $r = R/a$. Non-dimensional form of Eq. (2) can thus be represented as

$$\frac{d(\frac{R}{a})}{d(\omega t)} = \sin\phi - k$$

Or,

$$\frac{dr}{d\psi} = \sin\phi - k \tag{3}$$

In terms of the included angle $\phi$ and ratio of speed of dog to the speed of duck, $k = (v_1/v_2)$. The second kinematic Eq. (1) can be similarly re-written in non-dimensionalized form as

$$R\dot{\gamma} = v_2 \sin(\alpha_2 - \gamma) - 0 = v_2 \cos\phi$$

Which is,

$$R(\dot{\psi} + \dot{\phi}) = a\omega\cos\phi$$

Thus,

$$R(\omega + \dot{\phi}) = a\omega\cos\phi$$

Therefore,

$$\dot{\phi} = \frac{a\omega\cos\phi}{R} - \omega \tag{4}$$

Non-dimensionalizing with respect to $\psi$ and $r$ results in

$$\frac{d\phi}{d(\omega t)} = \frac{\cos\phi}{(\frac{R}{a})} - 1$$

Which is,

$$\frac{d\phi}{d\psi} = \frac{cos\phi}{r} - 1 \qquad (5)$$

Referring to Fig. 2, a possible engagement scenario is when speeds of the pursuer and target are equal (speed ratio, $k = 1$) and distance between them becomes zero ($r$ or $R = 0$). Interestingly, both these conditions including ($\phi = \frac{\pi}{2} rad$) together refer to an equilibrium state of the system (Eq. 2&4). How dynamics progresses when pursuer establishes contact with the target is analysed next in terms of equilibrium and stability as function of speed ratio.

## 3. Equilibrium, stability and phase plane analysis of the model

It is apparent that the model represented by Eqs. (2&4 or 3&5) is parametrized with the speed ratio and, in order to study this system, for each fixed value of speed ratio, one may resort to simulating the model from a given initial condition. Initial conditions (values of the states, $r$ and $\phi$, at time $t = 0$) can be arbitrarily or selectively chosen based on the region of interest. It is likely that choice of initial condition may lead to divergence of transients, abruptly terminating the computation. Dynamical system theory proves that initial conditions near stable equilibrium lead to convergence of trajectory while ones near unstable equilibrium lead to divergence of trajectory [11]. Thereby, concept of equilibrium and local stability becomes an important ingredient for the analysis of dynamical systems that are nonlinear. Bifurcation and continuation theory based methodology has proven to be very efficient for analysis of parameterized nonlinear dynamical systems. The methodology relies on computing equilibrium solution branches of nonlinear dynamical systems such as Eqs. (2&4 or 3&5) with local stability information of each equilibrium solution as one of the parameter of the system is varied. Numerical continuation tools available in public domain, for example, AUTO, COCO or MATCONT [12,13] are very popular and commonly used for this exercise. For our second order system, however, arriving at analytical solutions is plausible, which is presented next, and further, bifurcation analysis results are also presented.

Equilibrium of Eqs. (3&5) are:

$$\frac{dr}{d\psi} = 0 = sin\phi - k \Rightarrow sin\phi^* = k$$

and,

$$\frac{d\phi}{d\psi} = 0 = \frac{\cos\phi}{r} - 1 \Rightarrow \cos\phi^* = r^* = \pm\sqrt{1-k^2}.$$

Real solutions $r^*$ exist only for $0 \leq k \leq 1$.

Stability of equilibrium (*) solutions can be obtained by finding the eigenvalues of the Jacobian matrix

$$J = \begin{bmatrix} \frac{\partial f_1}{\partial r} & \frac{\partial f_1}{\partial \phi} \\ \frac{\partial f_2}{\partial r} & \frac{\partial f_2}{\partial \phi} \end{bmatrix} = \begin{bmatrix} 0 & \cos\phi^* \\ -\frac{\cos\phi^*}{r^{*2}} & -\frac{\sin\phi^*}{r^*} \end{bmatrix} \quad (6)$$

From properties of the matrix $J$, type of equilibrium solution based on stability can be studied.

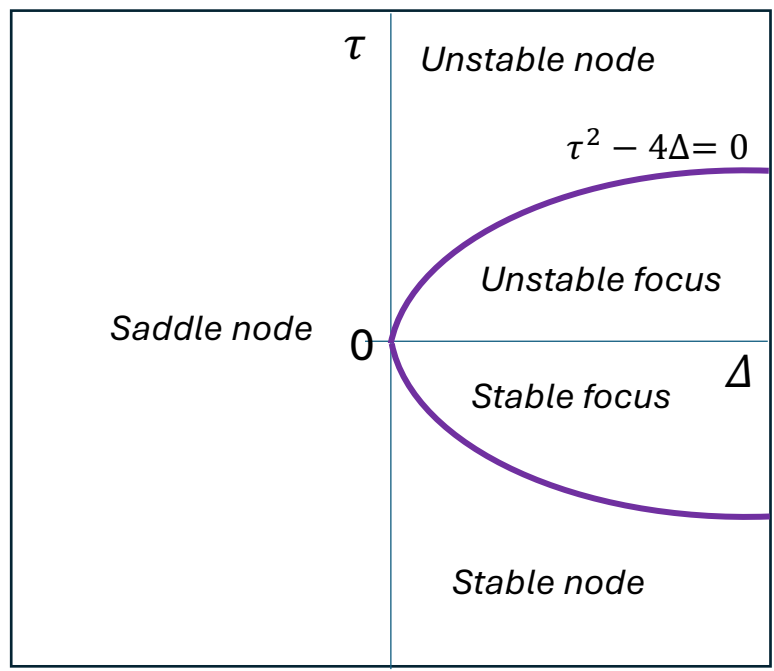

**Figure 3**: Eigenvalues and equilibrium type.

**Eigenvalues and stability**: For second order systems,

$$J = \begin{bmatrix} a & b \\ c & d \end{bmatrix} \quad (7)$$

Equation of characteristics is given by, $|\lambda I - J| = 0$, where $I$ is a 2 × 2 identity matrix and $\lambda$ are the eigenvalues. Equation of characteristics can be expanded as

$$\lambda^2 - (a+d)\lambda + (ad-bc) = 0 \tag{8}$$

which is,

$$\lambda^2 - \tau(J)\lambda + \Delta(J) = 0 \tag{9}$$

where $\tau$ is trace and $\Delta$ is determinant of $J$. Solution of Eq. (9) is,

$$\lambda = \frac{\tau \pm \sqrt{\tau^2 - 4\Delta}}{2}$$

Location of eigenvalues and type of equilibrium are presented in the schematic in Fig. 3.

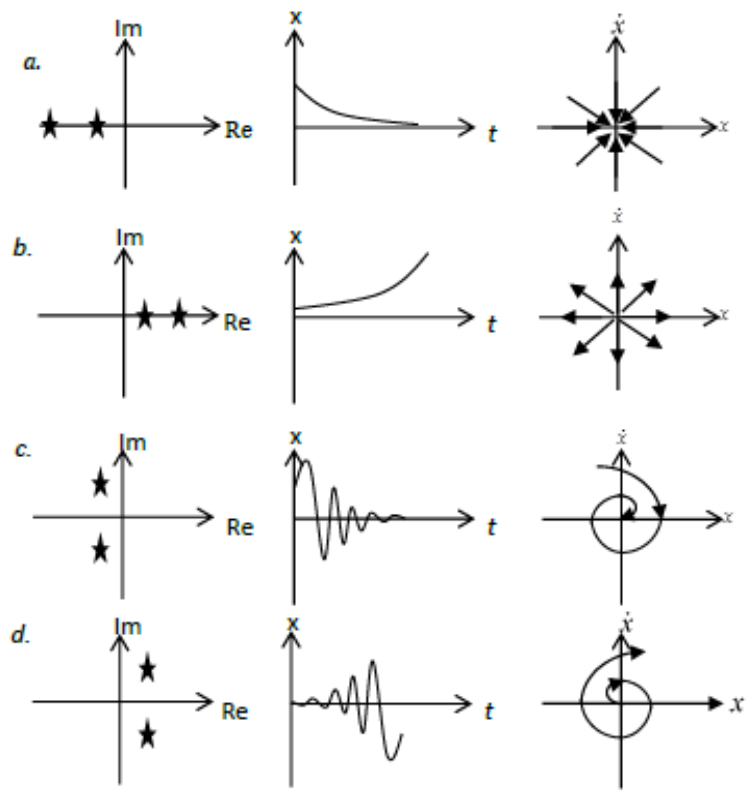

**Figure 4**: Eigenvalues and transients around an equilibrium state (Figure courtesy Ref.[14]).

For our system, Determinant $\Delta(J) = \cos^2\phi^* / r^{*2} = 1$; Trace $\tau(J) = -\frac{\sin\phi^*}{r^*} = -\frac{k}{\sqrt{1-k^2}}$

and $\tau^2 - 4\Delta = \frac{k^2}{(1-k^2)} - 4$. Trace $\tau$ is negative for all values of $k$ in the valid range $0 \leq k \leq 1$ for existence of equilibrium solutions. Since the determinant is constant and positive for all equilibrium states, the type of equilibrium states are determined by the sign of

$$\tau^2 - 4\Delta = \frac{k^2}{(1-k^2)} - 4$$

For all $\tau^2 - 4\Delta = \frac{k^2}{(1-k^2)} - 4 < 0$, i.e. $0 < k < \sqrt{4/5}$ (for positive $k$), the equilibrium states $(r^*, \phi^*)$ are stable focus (Fig. 4c). For $\tau^2 - 4\Delta = \frac{k^2}{(1-k^2)} - 4 > 0$, i.e. $\sqrt{4/5} < k < 1$, the equilibrium states $(r^*, \phi^*)$ are stable nodes (Fig. 4a). Thus, at critical $k = \sqrt{4/5}$, equilibrium type changes from a stable focus to a stable node, reflecting change in qualitative behaviour.

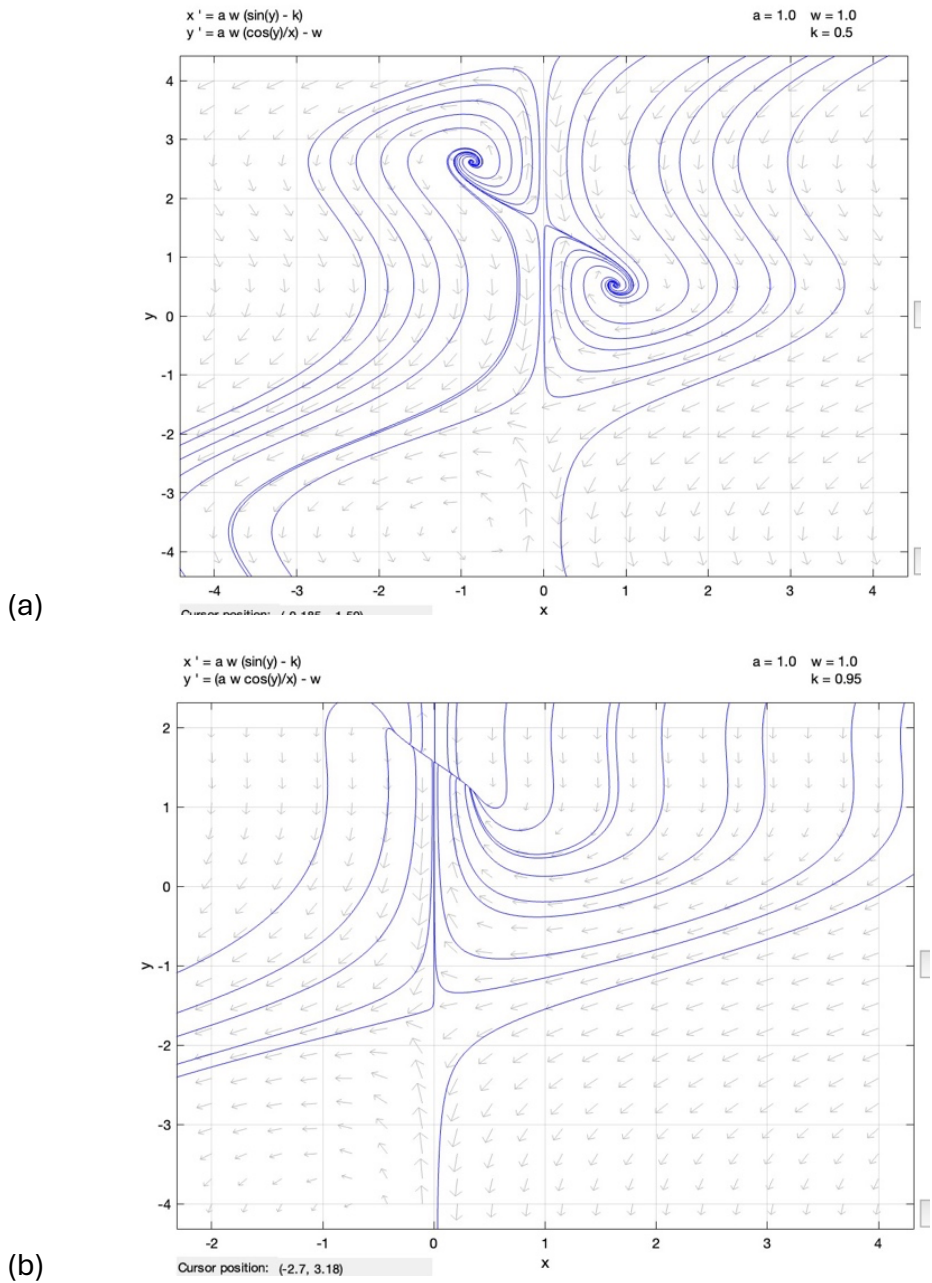

(a)

(b)

**Figure 5**: Phase portraits for two different values of speed ratios $k$.

What it means is, oscillatory transients of states around equilibrium state are now replaced with exponential transients as depicted in the phase portraits in Fig. 5. Figure 5 captures qualitative behaviour in real time (using original dynamic equations with explicit time derivatives) where radius of the circular path of the target and angular speed of the target are both assumed to be '1' without loss of generality. Phase portraits for two speed ratios in Fig. 5, plotted using 'pplane8.m' [15] from different initial conditions, corroborate well with the analysis presented earlier. First plot for speed ratio $k = 0.5 < 2/\sqrt{5}$ shows a stable focus at $(x, y) = (R, \phi) = (0.866m, 0.523rad)$ and second plot for speed ratio $k = 0.95 > 2/\sqrt{5}$ shows a stable node at $(x, y) = (R, \phi) = (0.312m, 1.253rad)$.

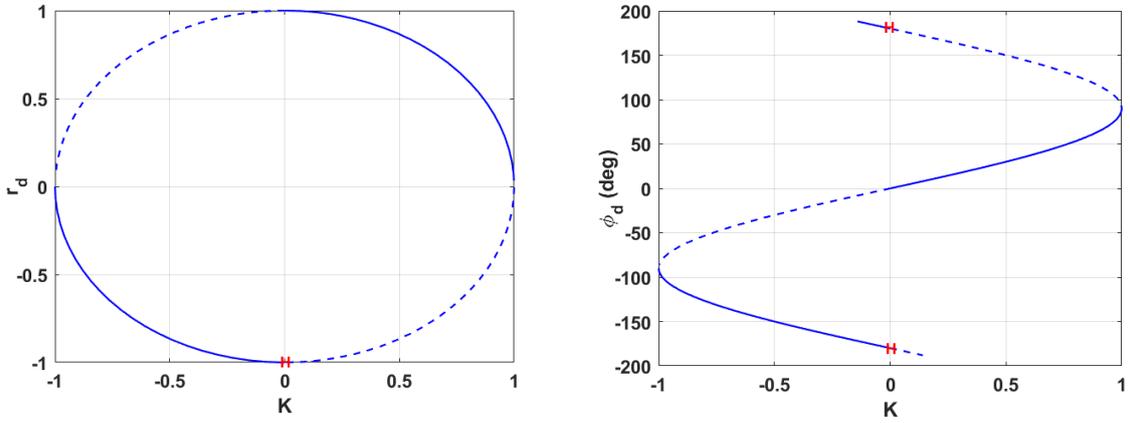

**Figure 6**: Bifurcation plots as function of speed ratio $k$ (solid lines: stable equilibrium, dashed lines: unstable equilibrium).

Bifurcation refers to a change in number of equilibrium states and/or their types with respect to a changing parameter. Bifurcation plots for the above model (Eqs. (3&5)) computed using a continuation algorithm are presented in Fig. 6. Matching speeds of target and pursuer (speed ratio $k = 1$) and orientation (angle $\phi = \pi/2$ rad) is the condition for engagement, which is also an equilibrium of the system as shown in Fig. 6. Solid lines are stable equilibrium states and dashed lines are unstable equilibrium states. Bifurcation results correlate well with the phase portraits in Fig. 5 at two values of the speed ratio. The plots also show solutions with negative speed ratio and negative distance, which is also output of the computation, that possibly can be related to a scenario of target moving in the clockwise direction (negative $k$ and $\phi$) and engagement. For the sake of current discussion, we focus on the solutions with physically realizable positive real values. Solutions to the 2-dimensional circular pursuit problem presented above are well known and match unambiguously with the solutions presented in [16].

Solutions presented above emphasize that to realise the engagement in practice the speed of the pursuer must be increased from any existing speed to atleast the speed of the target, whereby a mechanism must exist to accelerate the pursuer. For aerial vehicles that mechanism is thrust. Available force which is the difference of the available thrust and drag created (proportional to square of speed) must be large enough to meet the acceleration requirement which directly depends upon the radius of the circle and angular speed of the target in the circle. In another words, limits on available force automatically translates into maximum $(a\omega)$ that can be achieved in reality in an engagement. Some other considerations are, for example, how fast and what path would be optimal in an engagement. An attempt to answer these questions needs working with higher dimensional model and a computational approach to work with the higher order model, which is presented next, but still in a simplistic scenario for illustration. As stated above, reachability set of $(a\omega)$ in an engagement is directly dependent on the limits on the maximum available acceleration of the pursuer, thereby, modelling only pursuer suffices.

## 4. Model including pursuer dynamics

In the following, we consider simplest dynamic model for the pursuer, that introduces many parameters into an augmented model as we will see. Equation of the dynamics of pursuer limited by its own capability in terms of useful force can be written as

$$m\frac{dV}{dt} = T - D \qquad (10)$$

where $m$ is the mass of the pursuer and $v_1 = V$. $T$ is the thrust available and $D$ is the drag. The above equations can be non-dimensionalized as follows.

$$\frac{d(V/a\omega)}{d(\omega t)} = \frac{g(T-D)}{(mg)(a\omega^2)}$$

Considering pursuer to be an aerial vehicle, drag can be modelled as $D = \frac{1}{2}\rho V^2 S C_D$, where $\rho$ is the density of air, $S$ is a reference area (usually wing planform area), and $C_D$ is drag coefficient. Thus,

$$\frac{dk}{d\psi} = \frac{g(T-D)}{(mg)(a\omega^2)} = \frac{T}{ma\omega^2} - \frac{1}{2m}\rho a S C_D k^2 \tag{11}$$

Further, introducing throttle parameter, $\eta = (\frac{T}{T_m})$, where $T_m$ is the maximum thrust available, Eq. (11) can be re-written as,

$$\frac{dk}{d\psi} = \eta C_1 - C_2 k^2 \tag{12}$$

where $C_1 = (T_m/ma\omega^2)$ and $C_2 = (\rho a S C_D/2m)$ are constants.

Adding to the first two equations, governing equations of the dynamics are now represented by

$$\begin{aligned}\frac{dr}{d\psi} &= sin\phi - k = f_1 \\ \frac{d\phi}{d\psi} &= \frac{cos\phi}{r} - 1 = f_2 \\ \frac{dk}{d\psi} &= \eta C_1 - C_2 k^2 = f_3\end{aligned} \tag{13}$$

It is noticeable from the model of the pursuer (3rd Eq. in (13)) that amount of thrust available limits the speed of the pursuer and, therefore, also puts a limit on circle radius $a$ and angular speed $\omega$ of the target (reachable set) that can be eventually realized in an engagement. In the nonlinear model in Eq. (13), speed ratio $k$ is now a state of the system and throttle parameter $\eta$ is an independently varying parameter with respect to which engagement dynamics is to be investigated. Parameters for a real aircraft used in the following simulations are presented in Table 1.

Table 1: Parameter values used in computation

| Pursuer (Aircraft) parameters | Target parameters | Other parameters |
|---|---|---|
| Mass $m = 15118.35 kg$ | Radius of circle $a = 100m$ | Air density $\rho = 1.225\ kg/m^3$ |
| Maximum available thrust $T_m = 49817.6 N$ | Angular speed $\omega = 1 rad/s$ | Gravitational acceleration $g = 9.81\ m/s^2$ |
| Wing planform area $S = 37.16 m^2$ | | |
| Drag coefficient $C_D = 0.1423$ | | |

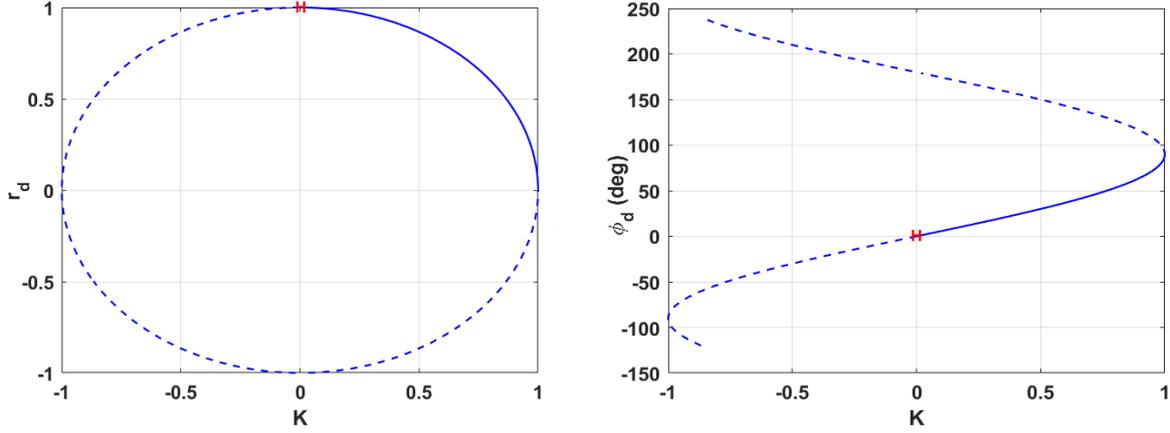

**Figure 7**: Bifurcation diagrams for the 3rd Order model as function of speed ratio $k$ (for $(a, \omega) = (100m, 1\,rad/s)$).

## 5. Results and Discussions

The 3rd order model includes many parameters (variables other than states), for example, radius of circle, angular speed of the target, physical parameters of the pursuer and above all the thrust that is limited by a maximum value. Speed ratio is no longer a parameter now as speed of the pursuer is governed by the dynamic equation (Eq. (13)). However, the condition for engagement still remains same as before, that is, speed of the pursuer must be equal to the speed of the target and the engagement occurs when $\phi = \pi/2$ rad. This requirement puts a demand on thrust which must be sufficient for an engagement or with the available thrust only targets with certain radius and angular speeds can be engaged. This is a vital information, and under no circumstances, engagement beyond available thrust can be realized. Bifurcation diagrams of 3rd order model are shown in Figs. 7&8. For comparison purposes, bifurcation diagram for $(r, \phi)$ are plotted in Fig. 7 as function of speed ratio $k$. Dependence of $r$ and $\phi$ upon speed ratio $k$ governed by first two equation in Eq. (13) remains same as before (Fig. 6), however with change in stability (solid lines: stable, dashed lines: unstable) for negative values of $(r, \phi, k)$. Inclusion of pursuer dynamics thus has effect on the stability of equilibrium solutions corresponding to negative $(r, \phi, k)$. Verification of change in stability and further insight into trajectories originating nearby any equilibrium solution on the bifurcation diagrams can be obtained by carrying out numerical time simulations of the Eq. (13). As stated before, however, we restrict our focus to physically realizable positive real equilibrium solutions. Bifurcation diagrams with respect to throttle (continuation) parameter $\eta$ are shown in Fig. 8.

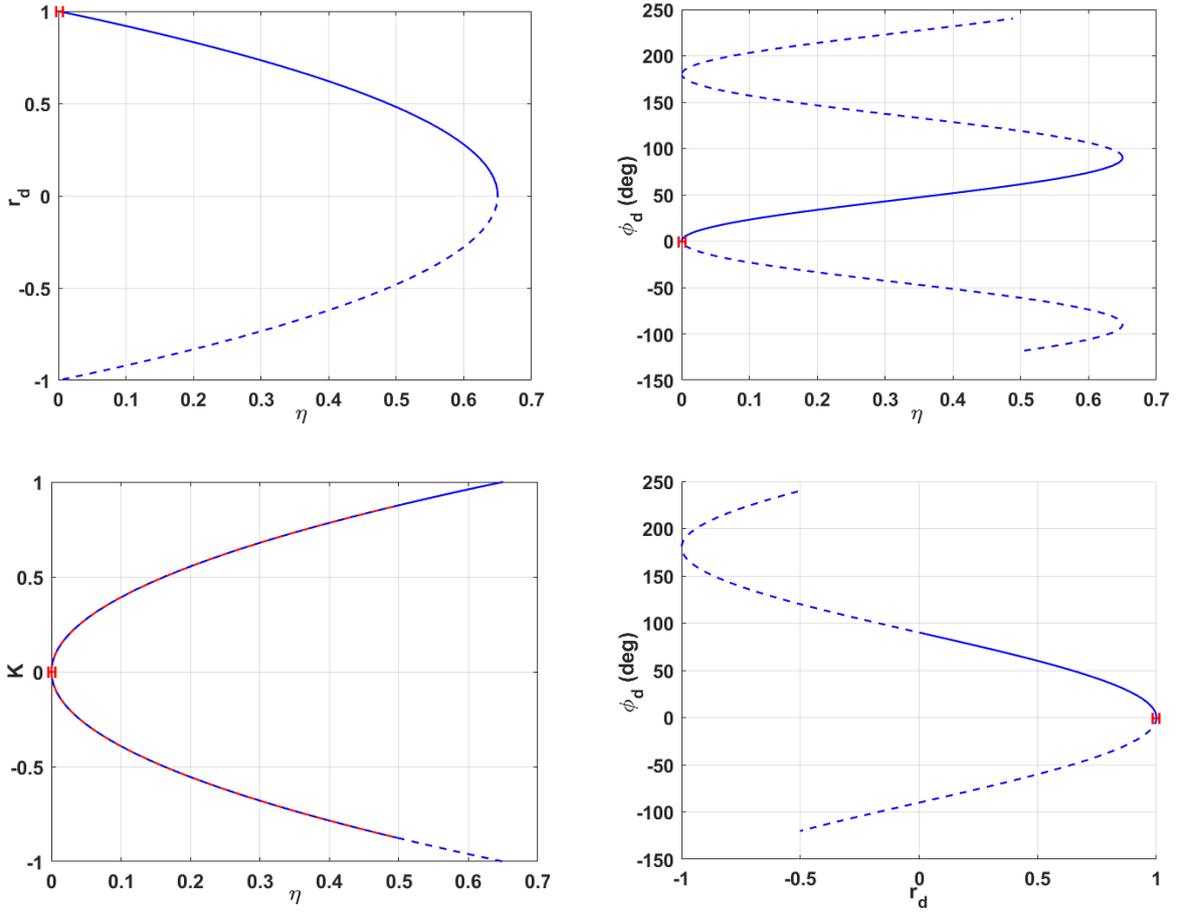

**Figure 8**: Bifurcation diagrams for the 3rd Order model as function of throttle parameter $\eta$ (for $(a, \omega) = (100m, 1rad/s)$).

Left bottom plot in Fig. 8 is variation of the speed ratio as function of the throttle parameter. Focusing on the positive values of $k$, it shows that $k$ reaches '1' at $\eta = 0.65$. Also, at $(k, \eta) = (1, 0.65)$, $(r, \phi) = (0, 90\text{deg})$ which is a condition for engagement. Starting from origin of the circle ($r = 1, \phi = 0, k = 0$), the pursuer speed increases till it reaches the terminal value, that is, the speed of the target going around in the circle and angle of engagement at $\phi = 90 deg$. Final value of thrust is about 65 percent of the maximum thrust ($\eta = 0.65$) required to engage, but all in steady state. Numerical value of $r_d (= R/a)$ goes from starting value of '1' to '0' as expected. The whole branch of equilibrium solutions in the positive range of speed ratio is stable and realizable.

Unlike in the previous case, for a 3rd order model, it is analytically difficult, if not impossible, to arrive at critical speed ratio at which type of equilibrium state changes. A continuation algorithm also outputs eigenvalues at each equilibrium state, which is presented in Table 2.

Table 2: Eigenvalues as function of speed ratio

| $\eta$ | $k$ | $\lambda_1$ | $\lambda_2$ | $\lambda_3$ |
|---|---|---|---|---|
| 0.487 | 0.865 | -0.03709 | -0.8645+0.5025i | -0.8645 - 0.5025i |
| 0.500 | 0.877 | -0.03757 | -0.9123+0.4092i | -0.9123 - 0.4092i |
| 0.518 | 0.892 | -0.03825 | -0.9910+0.1333i | -0.9910 - 0.1333i |
| 0.524 | 0.898 | -0.03848 | -0.8130 | -1.229 |
| 0.554 | 0.923 | -0.03957 | -0.5325 | -1.8771 |
| 0.649 | 0.999 | -0.02910 | -0.0428 | -30.435 |

From Table 2, a similar change in location of eigenvalues from a complex-conjugate pair (representing a stable focus) and a real to three real (representing a stable node) is noticed. This change in type of equilibrium state for the 3rd order model also happens at critical speed ratio of ($\frac{2}{\sqrt{5}} = 0.894$), as before, at throttle value of ~0.52.

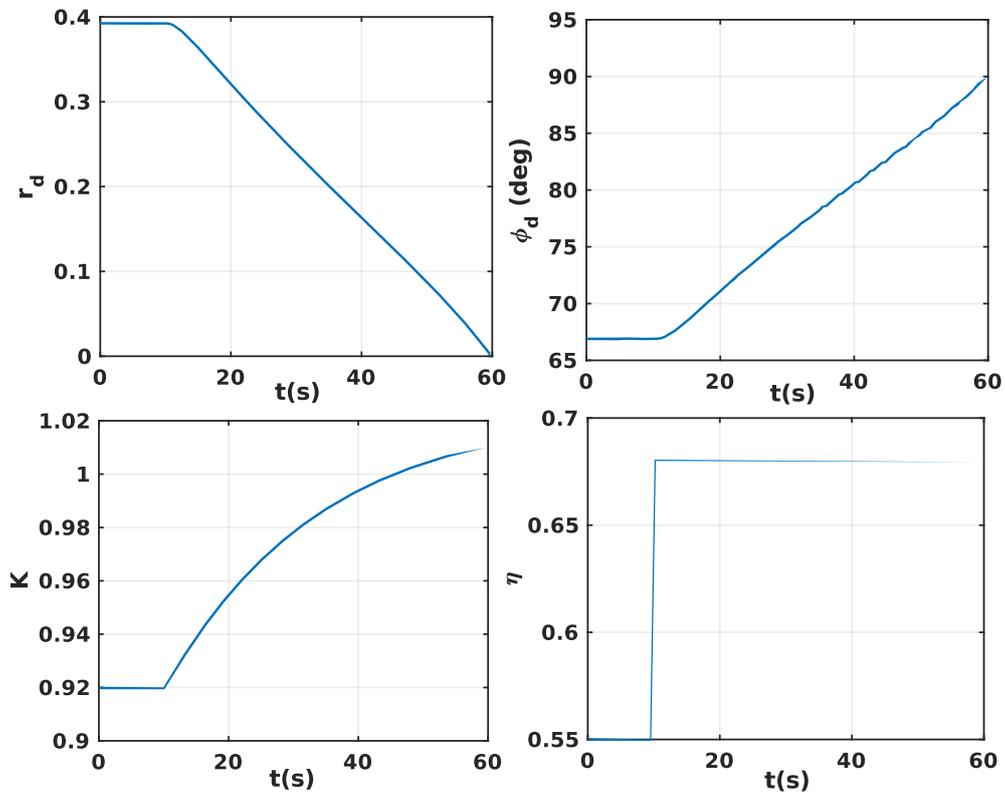

Figure 9 (a): Time simulation results for step input in throttle parameter.

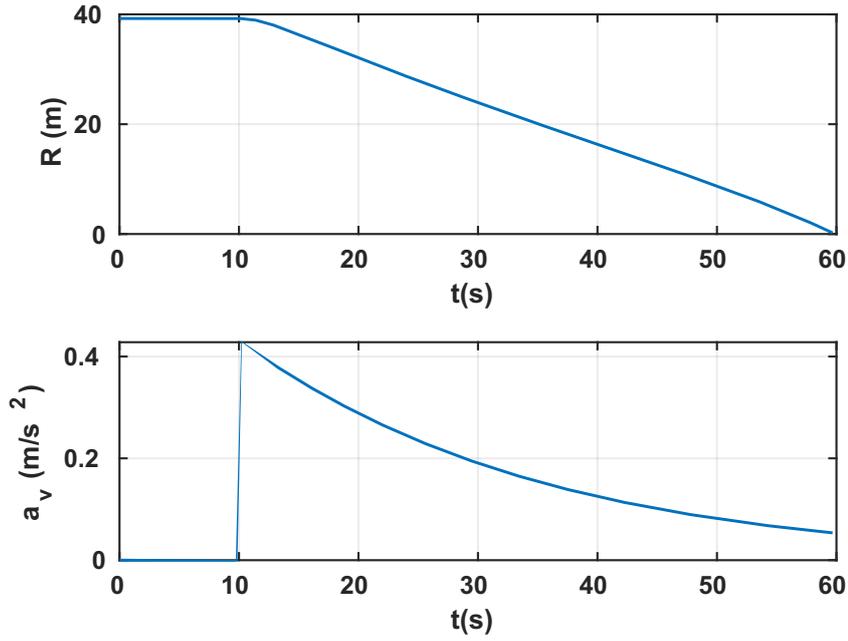

Figure 9(b): Time history of real distance and acceleration generated.

Bifurcation analysis results provide the minimum value of thrust that pursuer needs in order to engage with the target. Results also suggest that steady increase in thrust leading to steady increase in the speed of the pursuer will decrease the gap in speeds and eventually target can be engaged. However, in real scenario, it is expected that the pursuer engaging with the target will accelerate as soon as the target falls in line of sight and would engage at the earliest. As the transients are directly affected by location of eigenvalues and corresponding speed ratio, in the following simulations, advantage has been made of the fact that engagement occurs where the corresponding equilibrium state is a stable node. A scenario of engagement where thrust is increased in step (Fig. 9(a)) slightly beyond the minimum value ($\eta = 0.65$) applied at $t = 10s$ has been presented in Fig. 9. The plots in Fig. 9 show speed of pursuer increasing steadily beyond the speed of target (speed ratio $k > 1$) at which $(r, \phi) = (0, 90\text{deg})$ and an engagement is achieved. Alongside, time histories of pursuer's acceleration and real distance are also plotted in Fig. 9(b). Coming back to answer the question posed in problem description in Section 2, if the dog can ever catch the dog, the answer is 'YES' provided it can generate enough acceleration as demanded beyond the critical value of throttle ($\eta = 0.65$) obtained from the bifurcation analysis results. At any lesser value of throttle or developed acceleration, engagement can never happen. Simulation results in Fig. 10 where throttle is increased from a current value at $t = 0$ to ($\eta = 0.65$) show that distance between pursuer and target $R$ reduces nearly exponentially fast initially but approach speed slows down with increasing time resulting in $R \to 0$ only asymptotically.

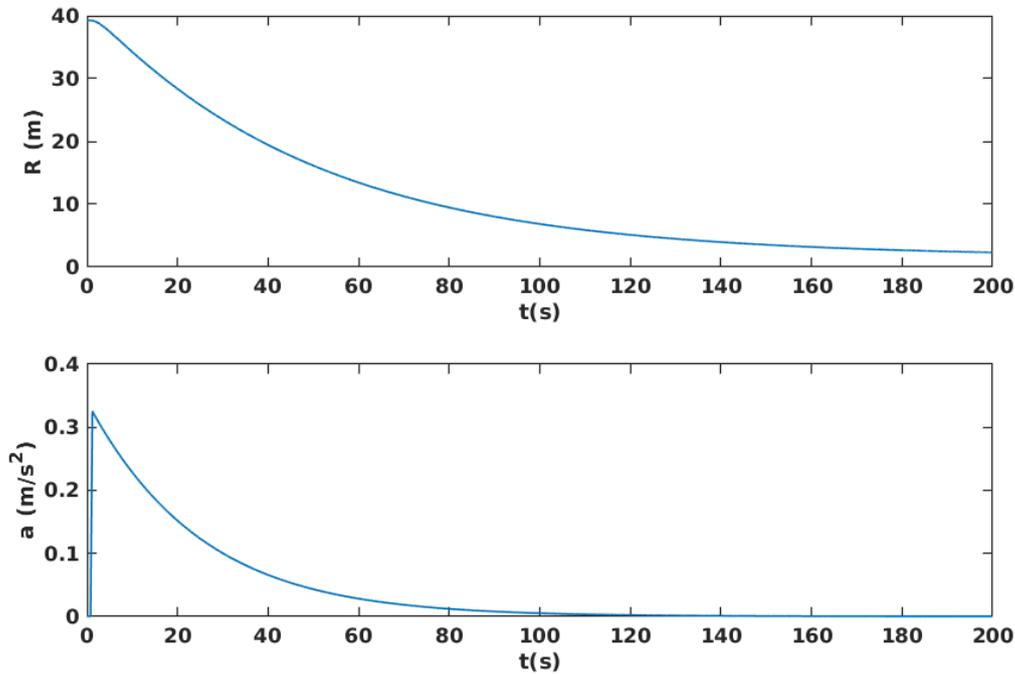

Figure 10: Time history of relative distance and acceleration for step increase in throttle to $\eta = 0.65$ at $t = 0$.

## 6. Conclusions

In this work, an elegant computational approach based on bifurcation theory and continuation methods is presented to study dynamics of a pursuit problem. A real world example is presented for a circular pursuit guidance of a pursuer towards a target going around in a circle. Equations of the engagement in a suitable form for bifurcation analysis are derived. Bifurcation analysis results followed by numerical simulations for two simplistic models, one without and the other with a dynamic model for pursuer, illustrating the concept are presented. Further analysis is required to determine the limits of $(a, \omega)$ that can be achieved with the thrust available. The approach being purely computational can be easily extended to include full order model of pursuer and/or target in entirety and guidance laws developed.